\documentclass[twoside]{article}
\usepackage{fleqn,espcrc2}

\usepackage{graphicx}
\usepackage[figuresright]{rotating}

\newcommand{\AmS}{{\protect\the\textfont2
  A\kern-.1667em\lower.5ex\hbox{M}\kern-.125emS}}

\font\quo=cmssqi8
\font\bsy=cmbsy10
\def\ub{\underbar}
\def\be{\begin{equation}}
\def\ee{\end{equation}}
\def\bea{\begin{eqnarray}}
\def\eea{\end{eqnarray}}

\title{KNO scaling 30 years later}

\author{S\'andor Hegyi\address{Particle Physics Department \\
        KFKI Research Institute for Particle and Nuclear Physics \\ 
        H-1525 Budapest 114, P.O. Box 49. Hungary}}

\begin{document}

\begin{abstract}
KNO scaling, i.e. 
the collapse of multiplicity distributions $P_n$ onto a universal 
scaling curve manifests when $P_n$ is expressed as the distribution 
of the standardized multiplicity $(n-c)/\lambda$ with $c$ and 
$\lambda$ being location and scale parameters governed by leading 
particle effects and the growth of average multiplicity. 
At very high energies, strong violation of KNO scaling behavior is 
observed ($p\bar p$) and expected to occur ($e^+e^-$). This challenges one 
to introduce novel, physically well motivated and preferably simple scaling
rules obeyed by high-energy data. One possibility what I find useful
and which satisfies the above requirements is the repetition of the original 
scaling prescription (shifting and rescaling) in Mellin space, that is, 
for the multiplicity moments' rank. This scaling principle will be 
discussed here, illustrating its capabilities both on model predictions and 
on real data.
\end{abstract}

\maketitle

\leftline{\quo Dedicated to Wolfram Kittel}
\leftline{\quo on the occasion of his 60th birthday}

\vspace{.5cm}

\section{THE EARLY YEARS}

\subsection{Polyakov and Koba-Nielsen-Olesen}

In the Millennium Year we celebrated the 30{\it th\/} anniversary of 
the famous Eq.~(1), the basic result of Polyakov and of
Koba, Nielsen and Olesen concerning the asymptotic behavior of the
multiplicity distributions~\cite{pol,kno1,kno2}. These authors put forward 
the hypothesis that at very high energies~$s$ the probability
distributions 
$P_n(s)$ of producing $n$ particles in a certain collision
process should exhibit the scaling (homogeneity) relation
\begin{equation}
        P_n(s)=
        {1\over\langle n(s)\rangle}\,\psi\!
        \left({n\over\langle n(s)\rangle}\right)
\end{equation}
as $s\to\infty$ with
$\langle n(s)\rangle$ being the average multiplicity of secondaries at 
collision energy~$s$. 
This so-called KNO scaling hypothesis asserts that if we rescale $P_n(s)$
measured at different energies via stretching (shrinking) the vertical
(horizontal) axes by $\langle n(s)\rangle$, these rescaled curves will
coincide with each other. That is, the multiplicity distributions
become simple rescaled copies of the universal function $\psi(z)$ depending
only on the scaled multiplicity $z=n/\langle n(s)\rangle$. In the picturesque
terminology of Stanley~\cite{sta} the rescaled data points $P_n(s)$ measured
at different energies~$s$ collapse onto the unique scaling curve $\psi(z)$.
After Koba~\cite{kno2}, let me illustrate this schematically:

\begin{center}
\includegraphics[height=4.8cm]{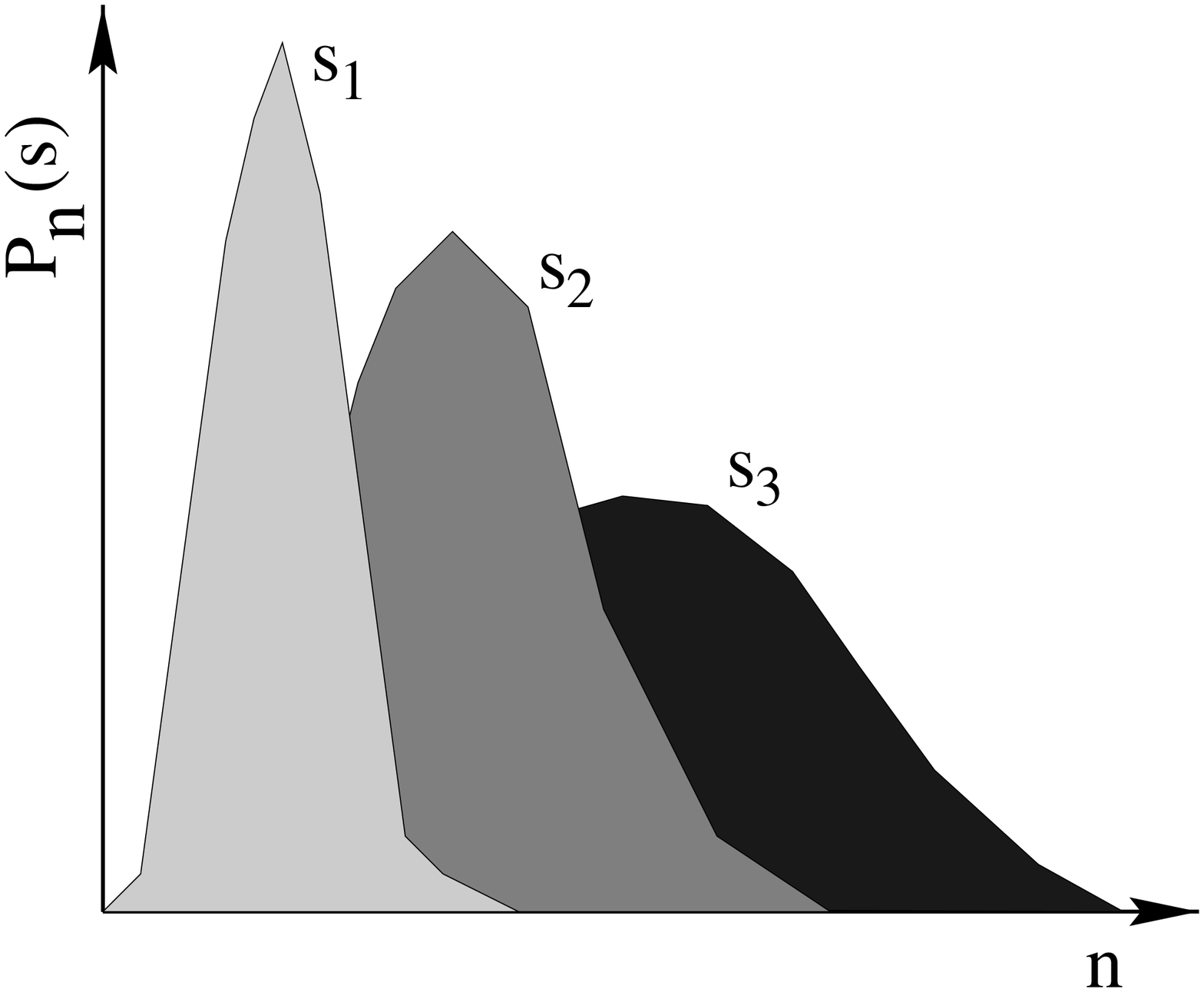}
\begin{eqnarray*}
\qquad\qquad
{\quad\quad\Big\Downarrow\quad\mbox{\fbox{\it rescaling\thinspace}}}
\end{eqnarray*}
\includegraphics[height=4.8cm]{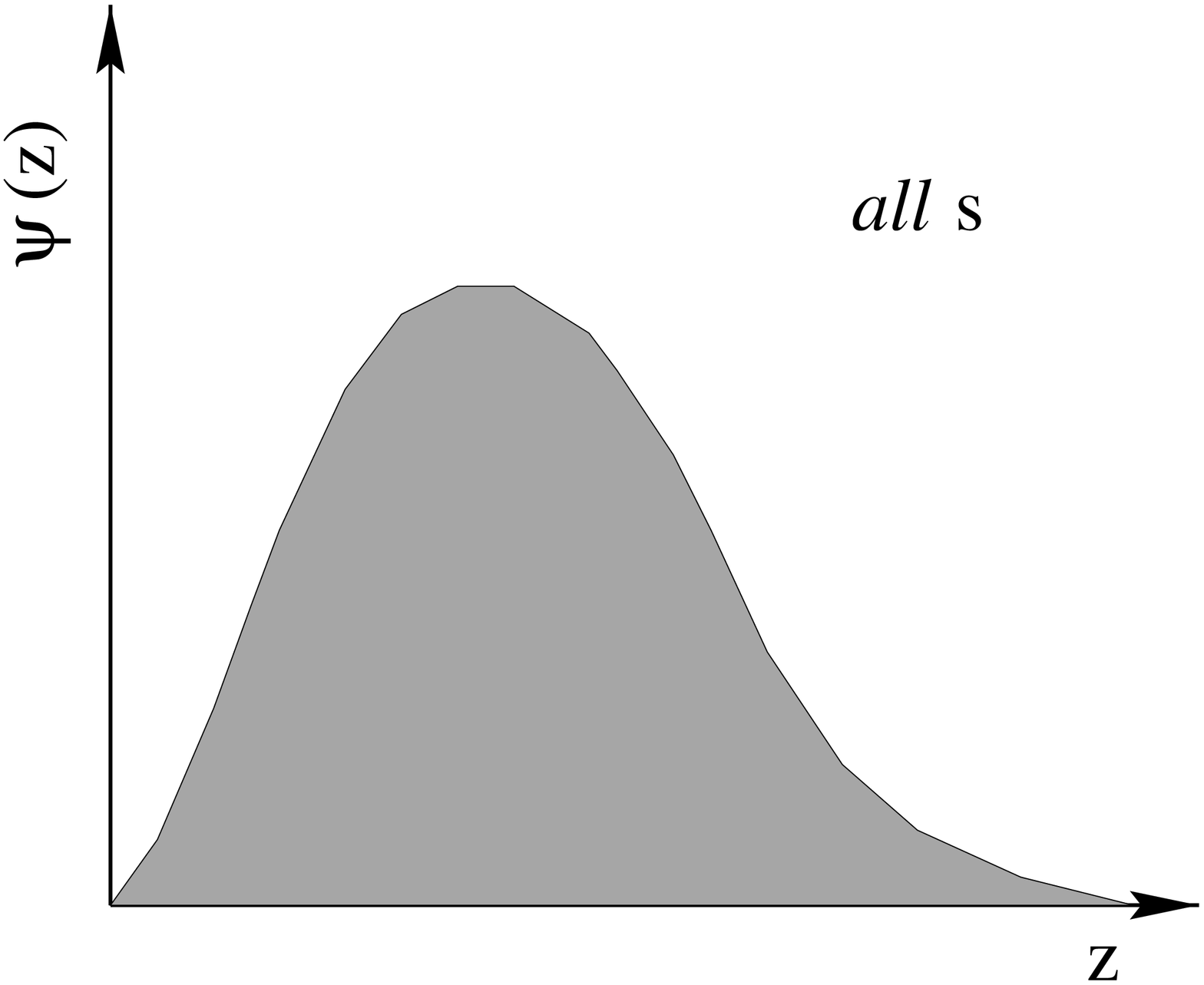}
\end{center}

\vspace{-.3cm}

The data collapsing behavior is a 
fundamental property of homogeneous functions.
According to Eq.~(1), at very high energies the multiplicity distributions 
$P_n(s)$ are homogeneous functions of degree $-1$ of $n$ and
$\langle n(s)\rangle$. Homogeneity rules play a central role in the
theory of critical phenomena~\cite{sta}. Near the critical point of a
physical system the thermodynamic functions exhibit homogeneous form
which implies the existence of a law of corresponding states: using a
suitably chosen scaling transformation it is possible to bring different 
states of the same system to coincidence and thus to compress many 
experimental or theoretical results into a compact form. The KNO 
scaling hypothesis Eq.~(1) is just such a law of corresponding states.

In the celebrated paper~\cite{kno1} Koba, Nielsen and Olesen showed 
that the validity
of Feynman scaling is a sufficient condition for Eq.~(1) to hold. It is 
worth mentioning however that the mathematical rigor of their derivation 
was criticized by several authors. Moreover, the precise finding of 
Ref.~\cite{kno1} was the energy independence of the moments
$C_q=\langle n^q(s)\rangle/\langle n(s)\rangle^q$ in the 
limit $s\to\infty$.
The scaling hypothesis Eq.~(1) was formulated with the additional
assumption that the moments $C_q=\int_0^\infty z^q\psi(z)\,dz$ determine
uniquely the scaling function $\psi(z)$. 
Polyakov arrived at the asymptotic multiplicity scaling law Eq.~(1)
two years earlier than Koba, Nielsen and Olesen by 
formulating a similarity hypothesis for strong 
interactions in $e^+e^-\to\mbox{\it hadrons\/}$ annihilation; see later.
Somehow this fundamental and elegant work was almost completely ignored 
in the early 70s and Eq.~(1) became known as the KNO scaling 
hypothesis in the high-energy physics community.

\subsection{Modifications}

The scaling relation Eq.~(1) can hold only approximately since
multiplicity~$n$ is a discrete random variable whose stretching or shrinking 
by a scale factor leaves the probabilities~$P_n$ unaltered. The proper 
meaning of Eq.~(1) is that with increasing collision energy~$s$ 
the discrete multiplicity distributions $P_n$ can be approximated
with increasing accuracy by a continuous probability density function
$f(x)$ via $P_n\approx f(x=n)$ (KNO prescription)
or by $P_n\approx\int_{x=n}^{x=n+1}f(x)\,dx$ (called KNO-G ``scaling'')
where, in the most popular modification of the original scaling rule, 
$f(x)$ has the generic Czy\.zewski-Rybicki form~\cite{cry,bla}
\be
	f(x)=\frac{1}{\lambda}\,\psi\left(\frac{x-c}{\lambda}\right)
\ee
with scale parameter $\lambda>0$ and location parameter $c$. 
Data collapsing behavior is observed if the only $s$-dependent parameters 
of the approximate shape function $f(x)$ are $c$ and $\lambda$, i.e. if 
$f(x)$ depends on collision energy only through a change of location 
and scale in~$x$. The variation of $\lambda$ with increasing~$s$
reflects the growth of average multiplicity, whereas the $s$-dependence 
of $c$ is usually associated with leading particle effects (so-called
AKNO or KNO-$\alpha$ scaling). Other physical explanations are also possible, 
see e.g. the contribution of Marek P\l oszajczak on $\Delta$-scaling.

\section{NEW SCALING PRINCIPLE}

Taking into account a possible location change (shift) in 
multiplicity rarely proved to be sufficient to restore the data
collapsing behavior of $P_n$ when the original scaling law Eq.~(1) is 
violated at very high energies (hence we use \hbox{$c=0$}). 
Nevertheless, the principle of scale {\it and\/} 
location change in the shape of multiplicity distributions can be 
extremely powerful through the following generalization of the KNO scaling
rule. Consider that the multiplicity distributions, more precisely the 
continuous probability densities $f(x)$ approximating $P_n$ generate
$s$-dependent shifting and rescaling in Mellin space.
The Mellin transform of a 
probability density $f(x)$ is defined by 
${\cal M}\{f(x);q\}=\int_0^\infty x^{q-1}f(x)\,dx$ and it provides the
$(q-1)${\it th\/} moment $\langle x^{q-1}\rangle$. Via the functional
relation
\begin{equation}
	{\cal M}\bigg\{x^r f\big(x^\mu\big);q\bigg\}=
	\frac{1}{\mu}\,{\cal M}\bigg\{f(x);{q+r\over\mu}\bigg\}
\end{equation}
one can introduce translation and dilatation in the moments' rank $q$ 
by performing the transformation $f(x)\to x^r f\big(x^\mu\big)$ of the 
probability densities approximating the shape of $P_n$. In the followings
I shall illustrate the utility of performing a location or scale change 
in Mellin space; for further details, see Refs.~\cite{hs1,hs2,hs3}.

\section{TRANSLATION IN {\bsy M}-SPACE}

\subsection{Motivation}

There are several possible dynamical explanations of
a shift in Mellin space.  The basic 
idea is simply the following. Recall that in Eq.~(1) the 
normalization of the KNO function reads
$
	\int_0^\infty\psi(z)\,dz=\int_0^\infty z\,\psi(z)\,dz=1,
$
that is, $\langle z\rangle=1$. Assume that violation of KNO scaling 
is observed and we measure $s$-dependent ``scaling'' functions
$\psi(z)$. In order to arrive at data collapsing behavior of $P_n$
the simplest possibility is to rescale the function
$\varphi(z)=z\,\psi(z)$ by its first moment given by $C_2$.
In terms of $P_n$ and the moments $\langle n^q\rangle$ 
this corresponds to rescaling 
the distribution $P_{n,1}=nP_n/\langle n\rangle$ by its mean 
$\langle n\rangle_1=\langle n^2\rangle/\langle n\rangle$. The distribution
$P_{n,1}$ is known in the mathematical literature
as the first {\it moment distribution\/} of $P_n$. If the
rescaling of $P_{n,1}$ by $\langle n\rangle_1$ is not sufficient to
arrive at data collapsing behavior,
one can try the same recipe for the higher-order moment distributions
$P_{n,r}=n^rP_n/\langle n^r\rangle$. The generalization is 
straightforward. Let me show you now an example when Eq.~(1) 
is clearly violated, but performing a shift in Mellin space can
restore the collapse of different $P_n$ curves onto a single 
scaling curve.

\subsection{Intermittency and multifractality}

The most obvious example is intermittency. As it turned out during the 
past 15 years, multiparticle production often yields scale-invariant
density fluctuations~\cite{bp,bop,ddk}. This can be observed
through the power-law scaling $C_q\propto\delta^{-\varphi_q}$ of the 
normalized moments $C_q=\langle n^q\rangle/\langle n\rangle^q$ 
of $P_n(\delta)$ as the bin-size~$\delta$ in phase-space is varied (for 
clarity, we neglect here the influence of low count rates).
Assuming that the fluctuations show monofractal structure,
the so-called intermittency exponents 
$\varphi_q$ are given by~$\varphi_q=\varphi_2(q-1)$ and the anomalous 
fractal dimensions $D_q=\varphi_q/(q-1)$ are \hbox{$q$-independent}, 
$D_q=D_2$. The normalized moments $C_q$ of $P_n(\delta)$ take the form
\begin{equation}
	C_q=A_q\,[C_2]^{\,q-1}\qquad\mbox{for}\quad q>2
\end{equation}
with coefficients $A_q$ independent of bin-size~$\delta$~\cite{ddk}. 
Of course Eq.~(1), with $s\to\delta$, is violated since we observe
$C_2\propto\delta^{-\varphi_2}$.

In the restoration of data collapsing behavior of $P_n$ for 
self-similar fluctuations the basic trick is the investigation 
of the higher-order moment distributions defined before. Their moments
are $\langle n^q\rangle_r=\langle n^{q+r}\rangle/\langle n^r\rangle$,
that is, the moments of the original $P_n$ are transformed out up to 
$r$th order via performing a shift in Mellin space, see
Eq.~(3). For $r=1$ the normalized moments of the 
first moment distribution $P_{n,1}$ are found to be
$
	C_{q,1}=C_{q+1}/[C_2]^{\,q}
$
in terms of the original $C_q$ and comparison to Eq.~(4) yields
$C_{q,1}=A_{q+1}$ for monofractal multiplicity fluctuations. Since the 
coefficients $A_q$ are independent of bin-size~$\delta$, we see
that monofractality yields not only 
{\it power-law scaling\/} of the normalized moments of $P_n$ but also 
{\it data collapsing\/} behavior of the first moment distributions $P_{n,1}$
measured at different resolution scales $\delta$. Increasing the 
(possibly fractional)
rank~$r$ of the moment distributions allows the restoration of data
collapsing behavior in the presence of increasing degree of multifractality
of scale-invariant multiplicity fluctuations~\cite{hs1,hs2}. The effect of 
low multiplicities (Poisson noise) can be taken into account via the study 
of {\it factorial\/} moment distributions 
$P_{n,r}=n^{[r]}P_n/\langle n^{[r]}\rangle$
and their factorial moments.

\subsection{Ginzburg-Landau phase transition}

In recent years considerable interest has been devoted to the
Ginzburg-Landau theory of the phase transition from quark-gluon plasma
to hadronic matter~\cite{hwa,hwb}. Instead of a strict power-law scaling
$F_q\propto\delta^{-\varphi_q}$ of the normalized factorial moments,
a so-called $F$-scaling behavior
$F_q\propto[F_2]^{\beta_q}$ is obtained with $\beta_q=(q-1)^\nu$.
For second-order transition $\nu=1.304$ and in case of a first-order
transition $\nu=1.45$. The $F$-scaling rule of the above form makes
difficult to identify the family of 
probability laws $P_n$ belonging to the model. But considering 
$P_{n,1}$ instead of $P_n$ the corresponding factorial moments 
turn out to be $F_{q,1}\propto[F_2]^{\beta_q}$ with $\beta_q=q^\nu-q$ and 
hence~\cite{hs1}
$$
        F_{q,1}\propto[F_{2,1}]^{\beta_q}\qquad\mbox{with}\quad
        \beta_q=\frac{q^\nu-q}{2^\nu-2}.
$$
This form of $F$-scaling is the familiar log-L\'evy law~\cite{lev} with
stability index $0<\nu\leq2$. Thus, via moment shifting, we succeeded 
in identifying $P_{n,1}$ for the Ginzburg-Landau formalism 
with a strict bound on the essential parameter~$\nu$ of the model.
The theoretically relevant values are within the allowed range.
Although this particular example is not related directly to data collapsing
behavior of multiplicity distributions, it illustrates the utility of the 
moment distributions $P_{n,r}\propto n^rP_n$ which
correspond to translation in Mellin space.

\section{RESCALING IN {\bsy M}-SPACE}

\subsection{Motivation}

The most important reason of a possible change of scale in
Mellin space comes from QCD. In higher-order pQCD calculations,
allowing a more precise account of energy conservation in the course of
multiple parton splittings, the natural variable of the multiplicity 
moments is the {\it rescaled\/} rank $q\gamma$ instead of rank $q$ 
itself~\cite{dok,och,dre} with
$\gamma(\alpha_s)$ being the multiplicity anomalous dimension of QCD.
Because of the running of the strong coupling constant $\alpha_s$,
it is inevitable to adjust an energy dependent scale factor in 
Mellin space if one wants to arrive at data collapsing behavior
of the multiplicity distributions $P_n(s)$.

\subsection{Running coupling}

In the very first paper predicting the multiplicity scaling law Eq.~(1),
Polyakov constructed a model for the short-distance behavior of strong
interactions based on the hypothesis of asymptotic scale- and conformal
invariance~\cite{pol}. When applied to the
process of $e^+e^-$ annihilation to hadrons at asymptotically high 
collision energies, the model predicts that multihadron production 
goes in a cascade way. First, a few heavy virtual objects, called jets, are
formed. These are repeatedly diminished in a branching process 
until the masses of the subsequent generation jets become at some 
very late stage  comparable to the hadronic masses. This cascading mechanism
gives rise to multiplicity distributions $P_n(s)$ satisfying Eq.~(1) and
the scaling function $\psi(z)$ behaves as
$
	\psi(z)\propto a(z)\,\exp(-z^\mu)
$
with $\mu>1$ and $a(z)$ being a monomial. 
Up to a scale factor, $\psi(z)$ is a negative
binomial type scaling function, that is, a gamma distribution~\cite{cs} 
in the rescaled and power-transformed multiplicity~$z^\mu$.

It is worth emphasizing that the above result was achieved years earlier
than the emergence of quantum chromodynamics. Interestingly, the predictions
of QCD obey some important similarities to the findings of Polyakov:

{\it 1)\/} 
Hadronic jets produced in hard processes are expected in QCD to be
self-similar objects composed of subsequent 
parton branching decays~\cite{kon}. A highly
virtual quark or gluon decays into secondaries which are less off-shell and
less energetic. Each of these intermediate decay products splits into novel
ones. The subsequent decay chain steps form a scale-invariant branching 
process which continues until the virtual mass of the quanta becomes
approximately 1~GeV.

{\it 2)\/} 
The fragmentation of partons in QCD leads to the typical characteristics of
branching processes, in particular, to long-range correlations and to the
validity of KNO scaling~\cite{bas}. Further, it was shown~\cite{tes} that
KNO scaling in $e^+e^-$ annihilation is the consequence of the 
scale-invariance
(asymptotic freedom) in the presence of gluon self-coupling. It is thus
the manifestation of the essence of the non-abelian gauge theory of
strong interactions.

{\it 3)\/} 
Taking into account the higher pQCD effects responsible for
energy-momentum conservation in parton jets, the KNO scaling function
in $e^+e^-$ annihilation was shown~\cite{dok} to be identical to 
Polyakov's form:
\be
	\psi(z)={\cal N}z^{\mu k-1}\exp\big(-[Dz]^\mu\big)
\ee
where $k=3/2$, ${\cal N}=\mu D^{\mu k}/\Gamma(k)$, 
$\mu=(1-\gamma)^{-1}\approx5/3$ and $D$ is a scale parameter
depending on $\gamma$, with $\gamma$ being the QCD anomalous dimension,
$\gamma\approx0.4$ at LEP1. Hence, the
conservation laws strongly reduce the width of the KNO scaling function 
as compared to the exponential fall-off obtained by lower-order pQCD 
calculations. The experimental data at $\sqrt s=91.2$~GeV (dots) confirm
the prediction (curve) as illustrated in the figure~a) below
(with the exception that $k$ turns out to be larger, $k\approx5$).

The crucial difference between the QCD prediction and Polyakov's result
lies in the fact that the scaling function $\psi(z)$  given by the 
Polyakov-Dokshitzer form Eq.~(5) {\it depends\/} on 
collision energy~$s$ in QCD. That is, QCD predicts violation of 
KNO scaling in $e^+e^-$ annihilation which becomes clearly visible 
at higher energies. This effect is due to the running
of the strong coupling constant $\alpha_s$, hence       
$\gamma\to0$ as $s\to\infty$, i.e., $\mu\to1$
asymptotically~\cite{dok}. Therefore the tail of $\psi(z)$ widens
with increasing~$s$ which gives rise to the KNO scaling violation pattern
shown in the figure~b) below. Asymptotically the exponential fall-off 
predicted by the double logarithmic approximation is recovered (solid curve) 
i.e. the negative binomial type scaling form of $P_n(s)$.

\vspace{-.2cm}

\begin{center}
\hbox{\hspace{-.2cm}\includegraphics[width=7.2cm]{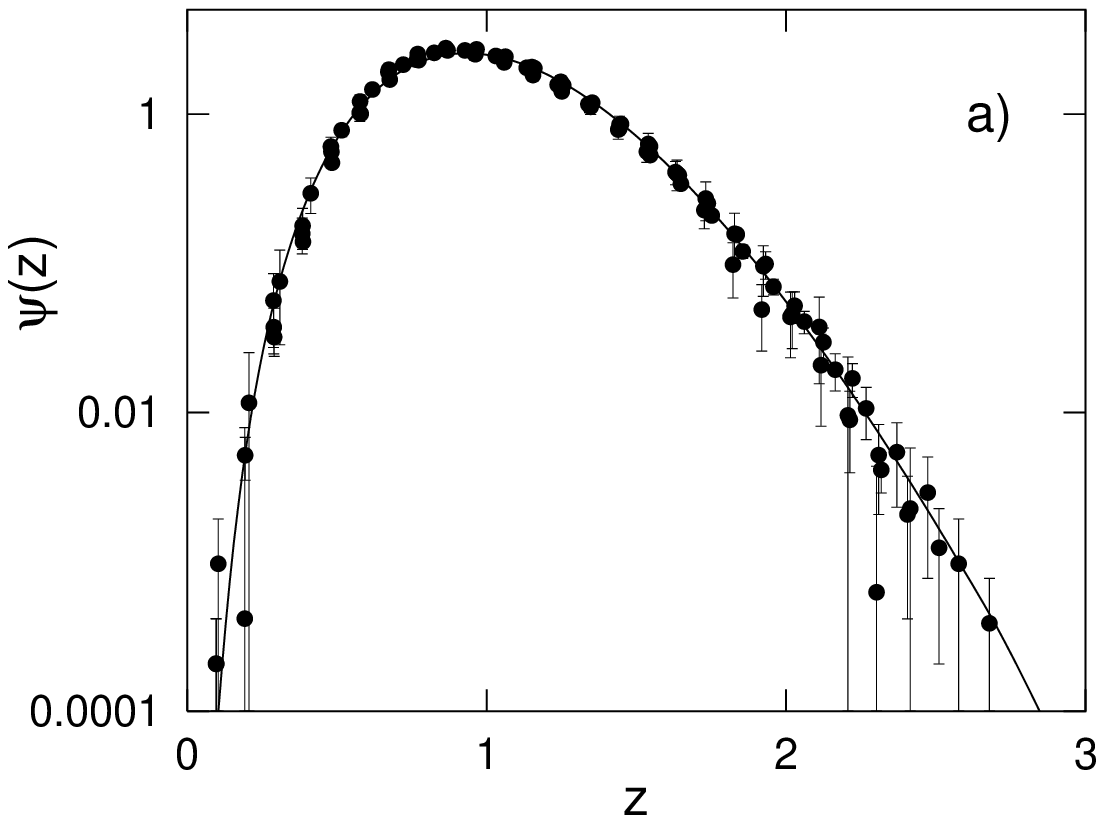}}
\hbox{\hspace{-.2cm}\includegraphics[width=7.2cm]{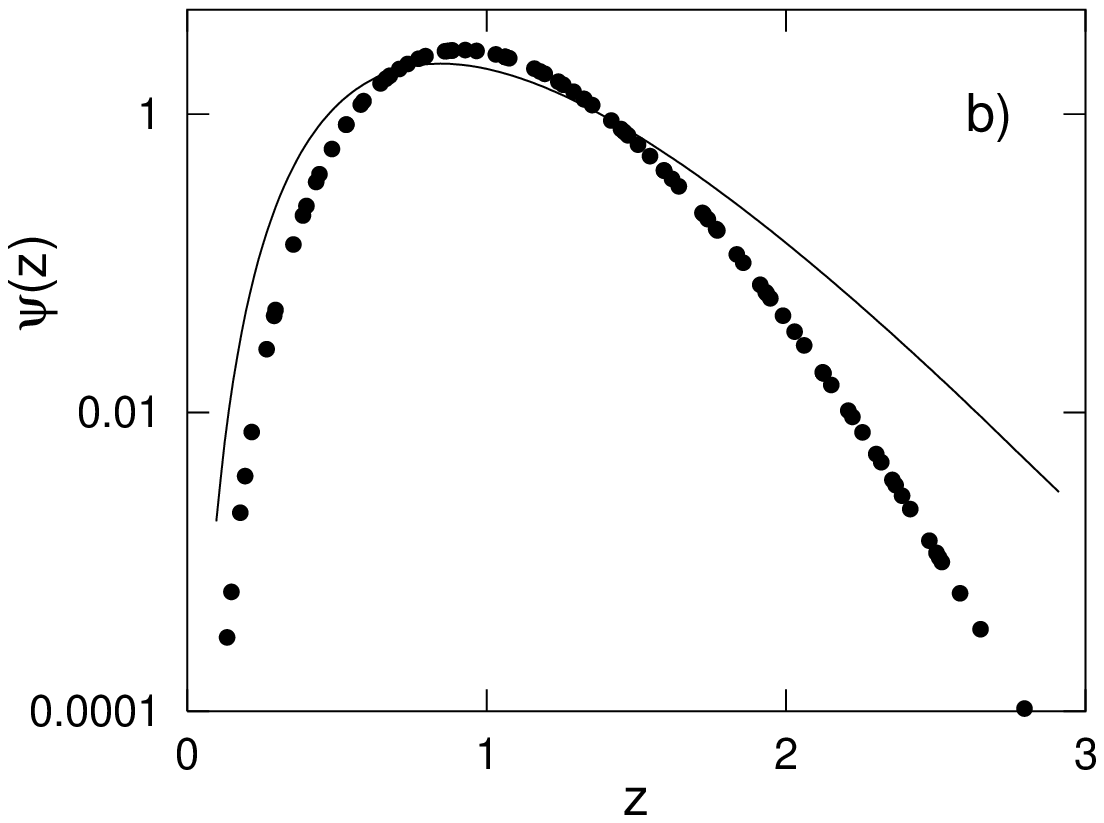}}
\hbox{\hspace{-.2cm}\includegraphics[width=7.2cm]{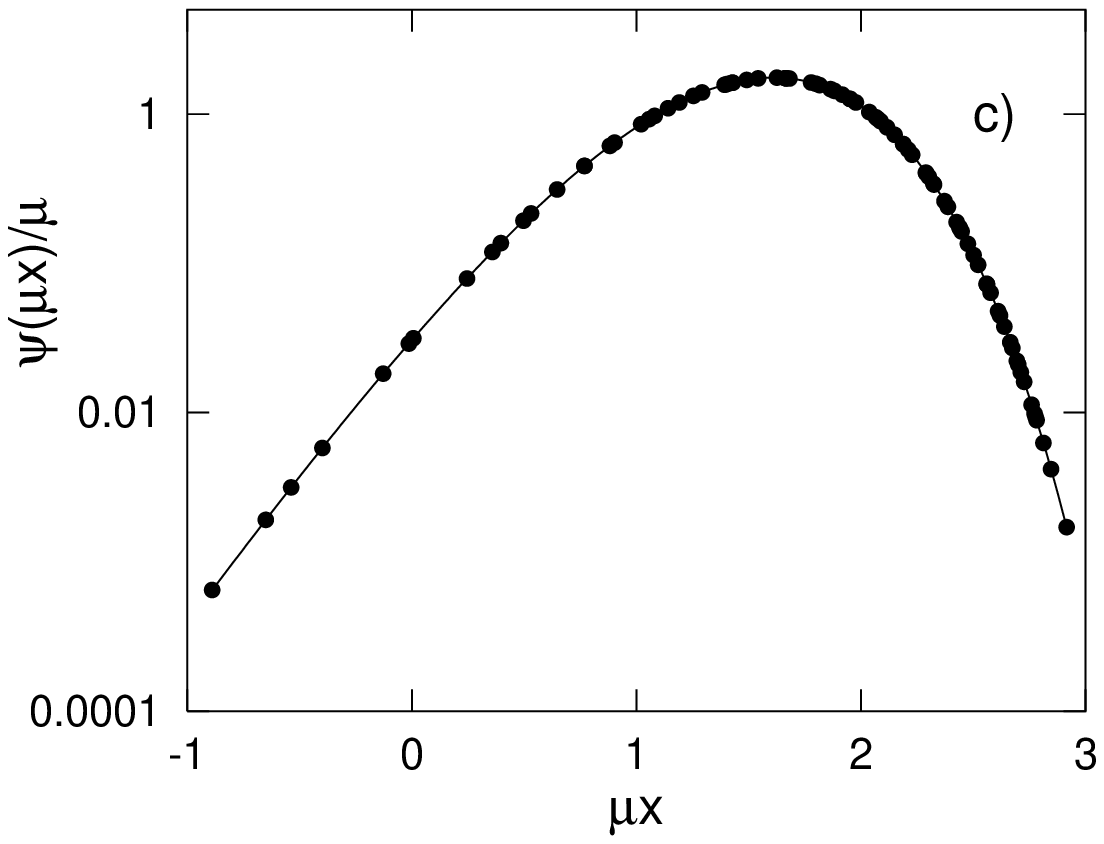}}
\end{center}

\vspace{-.7cm}

Thus we have learned that the pre-QCD and QCD-based descriptions of $e^+e^-$
annihilation to hadrons make essential use of self-similar branching
processes, further, both predict the rescaled and
power-transformed multiplicity variable~$z^\mu$ 
to be gamma-distributed. But in the pre-QCD approach
the exponent $\mu$ is independent of collision energy~$s$ and
KNO scaling holds valid, as expected for scale-invariant branching with 
{\it fixed coupling\/}. On the contrary, the QCD-based calculations
predict violation of KNO scaling since $\mu$ varies with collision energy~$s$, 
due to the {\it running\/} of $\alpha_s$. Note however that
in the latter case data collapsing can be restored in a simple way using
{\it logarithmic\/} scaling variable. For the Polyakov-Dokshitzer form
of $\psi(z)$ given by (5) we have
$$
	\psi(x)=\mu\exp\big(k\mu x-e^{\mu x}\big)/\Gamma(k),
	\quad x=\ln(Dz).
$$
Since only the exponent $\mu$ and scale parameter $D$ of Eq.~(5) are 
expected to depend on collision energy~$s$ through the variation of 
$\gamma(\alpha_s)$, KNO scaling violated by 
QCD effects is recovered by plotting $\mu^{-1}\psi(\mu x)$ as displayed in 
the figure~c). The scale change in logarithmic multiplicity is 
governed by the QCD multiplicity anomalous dimension. This particular
type of data collapsing of the multiplicity distributions $P_n(s)$ 
is called log-KNO scaling~\cite{hs2,hs3} since we have the behavior
of type Eq.~(2) except that distributions of the logarithmic 
multiplicity satisfy it.

\subsection{Multiplicative cascades}

It may turn out that the log-KNO scaling law
has ubiquitous appearance in multiparticle dynamics.
Random multiplicative cascades play a distinguished role in the dynamics
underlying multihadron production, both in soft and hard processes.
The multiplicative cascade models like for example the
$\alpha$-model and $p$-model are in the focus of interest since the
pioneering work of Bia\l as and Peschanski~\cite{bp}, see also~\cite{ddk}
for a review. Recently, Frisch and Sornette developed a theory of 
extreme deviations~\cite{frs} which predicts for multiplication
of random variables the generic presence of stretched exponential
distributions $f(x)\propto\exp(-\lambda x^\mu)$ with $\mu<1$. 
To be specific, consider the product
$
	X_N=m_1m_2\cdots m_N
$
of independent random variables $m_i$ distributed identically according to
the probability density $p(m_i)$. The above product can arise in a cascade 
process if primary entities (e.g. particle density) of initial size $s_0$
are repeatedly diminished in random proportions, so that after $N$ cascade
steps we have the product \hbox{$s_N=s_0m_1m_2\cdots m_N$}.
\begin{center} 
\includegraphics[height=4cm]{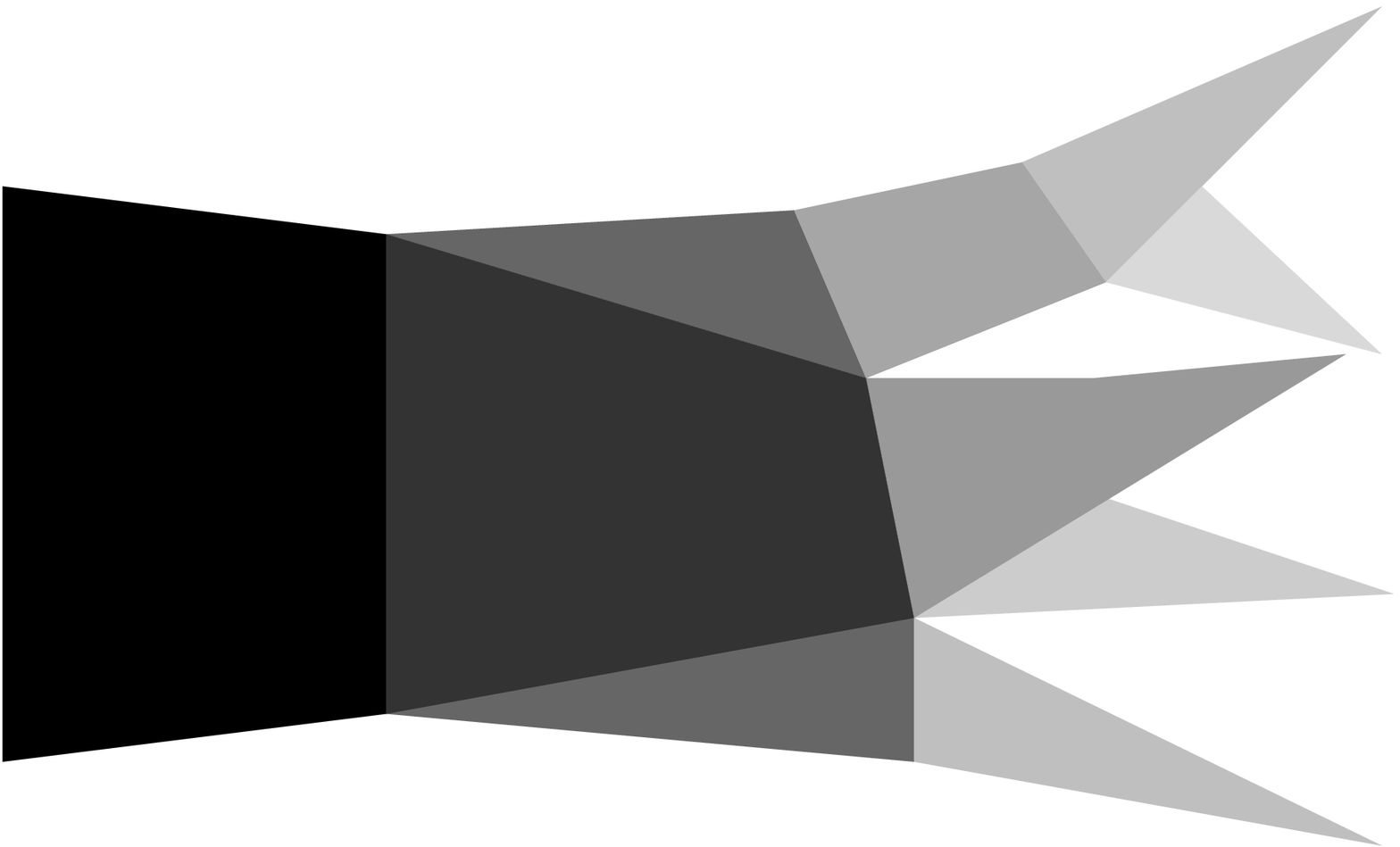}
\end{center}
In ref.~\cite{frs} it is shown that under
mild regularity conditions the distribution of the product variable $X_N$ 
exhibits the asymptotic behavior
\be
	{\cal P}_N(X)\sim\big[p\big(X^{1/N}\big)\big]^N
	\qquad\mbox{for}\quad X\to\infty
\ee
and for finite~$N$. Frisch and Sornette call attention to the intuitive
interpretation of (6): the tail of ${\cal P}_N(X)$ is controlled by
the realizations where all terms in the product are of the same order
hence ${\cal P}_N(X)$ is, to leading order, just the product of the $N$
densities $p$, each having the common argument $X^{1/N}$. When $p(x)$ is
chosen to be $\propto\exp(-\lambda x^\alpha)$ with $\alpha>0$,
then Eq.~(6) leads to stretched exponential tails
$\propto\exp(-\lambda Nx^{\alpha/N})$ for large $N$, with stretching
exponent $\alpha/N<1$. 
Expecting the cascade depth $N$ to be an increasing function of collision
energy $s$, Eq.~(1) breaks down but data collapsing can be recovered in
the log-KNO manner discussed previously.

\section{SCALING AT TEV ENERGIES}

The most exciting and challenging task in developing novel scaling 
relations is testing them on real data. In case of multiplicity 
distributions the KNO scaling laws Eqs.~(1-2) are known to be strongly
violated above ISR energies. This violation is most visible for the
multiplicity data measured by the E735 Collaboration~\cite{e735} see 
also~\cite{wal} for more details. In the followings I provide a very brief
summary of some preliminary results of a study of the E735 data.

\subsection{Analysis of E735 multiplicities}

The E735 Collaboration recently published the full phase-space 
multiplicity distributions in $pp$ and $p\bar p$ collisions
at c.m. energies $\sqrt s=$ 300, 546, 1000 and 1800 GeV 
at Tevatron~\cite{e735}. The
 $P_n$ curves corresponding to $\sqrt s=$ 300 and 1800 GeV are 
displayed in the following two figures. 
\begin{center}
\hbox{\hspace{-.4cm}\includegraphics[width=8cm]{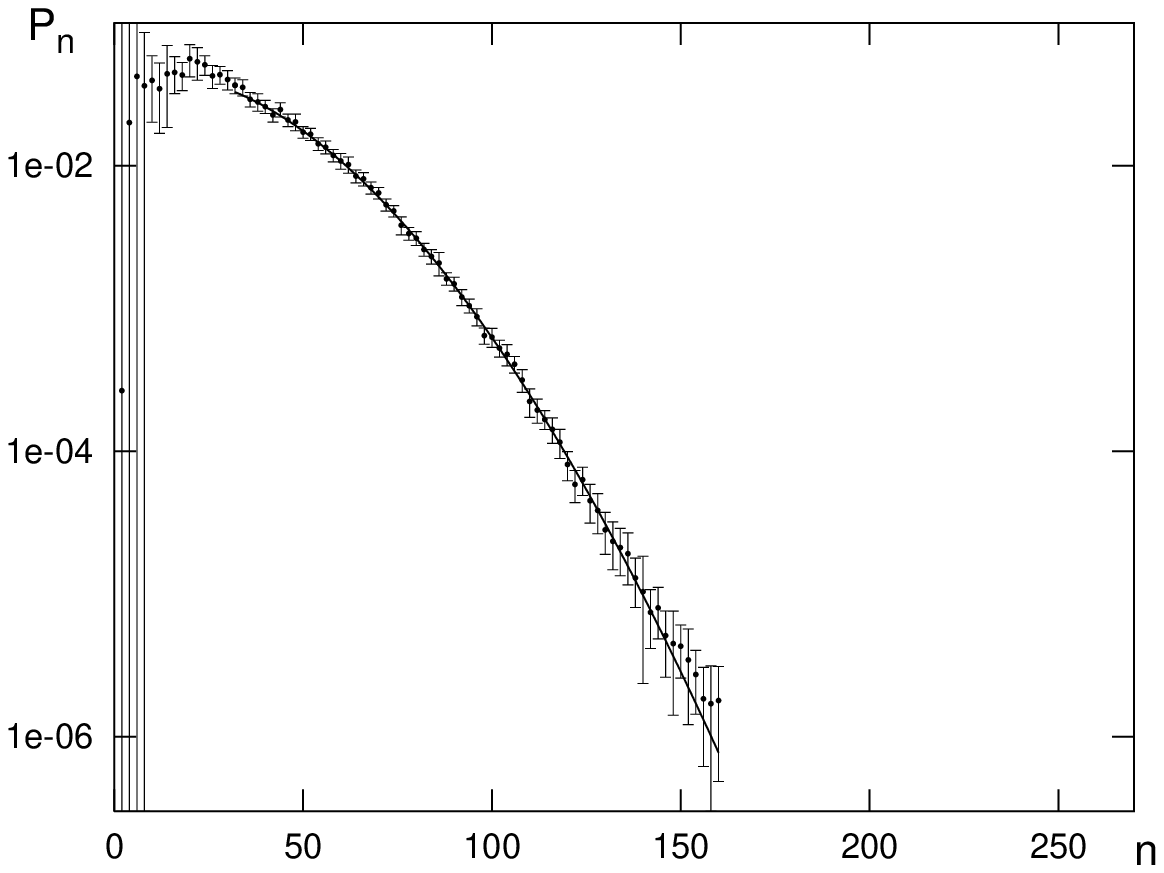}}
\hbox{\hspace{-.4cm}\includegraphics[width=8cm]{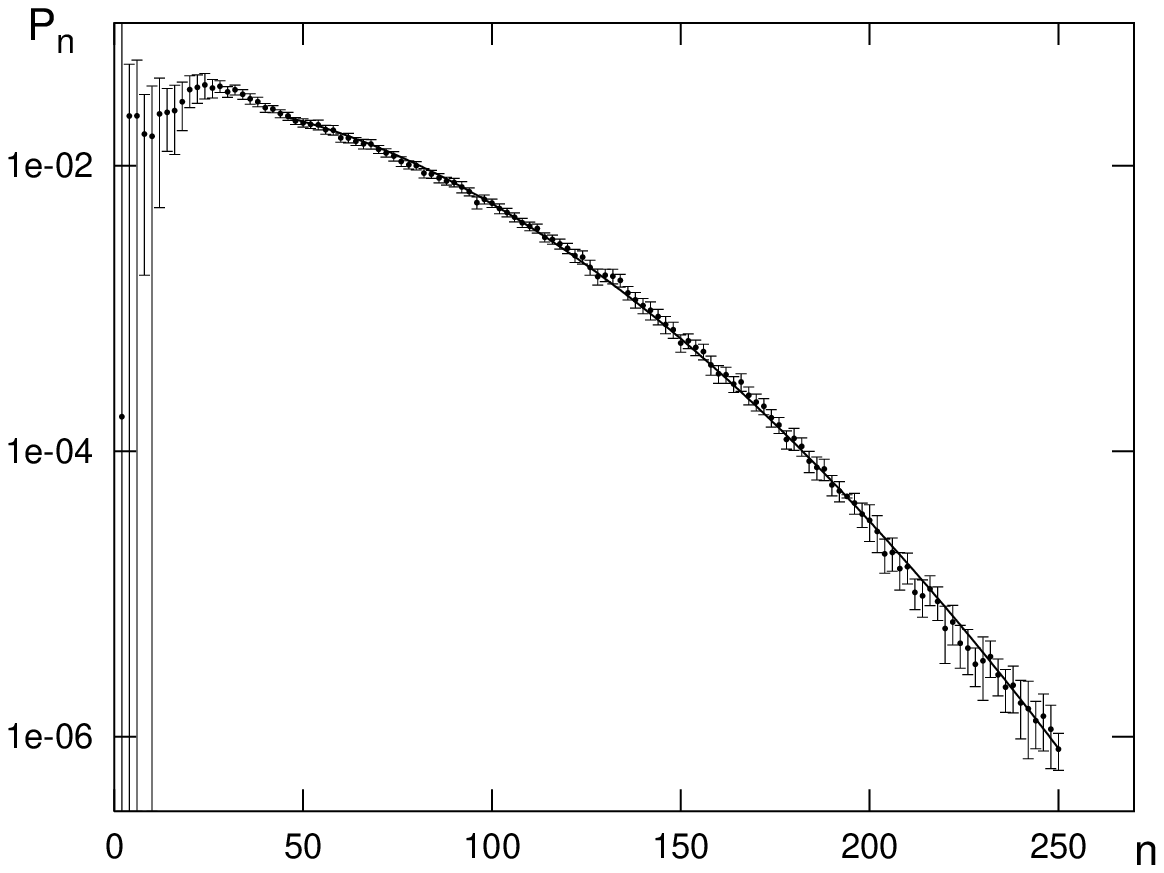}}
\end{center}

It is apparent that bimodal shape of the $P_n$ curves arises at
TeV energies, obeying a (not too pronounced) shoulder structure like at SPS.
It is argued~\cite{e735,wal}
that the low-multiplicity regimes are affected mainly by single parton 
collisions and exhibit KNO scaling, 
whereas the \hbox{large-$n$} tail of the distributions is
influenced more heavily by double parton interactions and violate 
KNO scaling substantially. Another possible explanation of the
observed $P_n$ shape is the weighted superposition of a soft and
semihard component, the latter one being dominated by minijet production.
These components are responsible of the KNO scaling and KNO-violating
regimes of $P_n$~\cite{gu1,gu2}.

The solid curves in the previous two plots represent the Polyakov-Dokshitzer 
parametrization Eq.~(5) fitted to the KNO-violating $z\geq1$
component of the E735 data. The quality of fits is reasonably good 
(for all data sets) as can be depicted from the figures.
The fits correspond to $k=\frac{1}{2}$ kept fixed and $\mu$ decreasing from
$\mu\sim2.2$ to $\mu\sim1.7$ as one goes from 
$\sqrt s=$ 300 GeV to 1800 GeV. Our findings
indicate that the KNO violating component of the E735 multiplicity
distributions can be collapsed onto a single scaling curve 
in the log-KNO manner discussed previously. It is illustrated below.
\begin{center}
\hbox{\hspace{-.4cm}\includegraphics[width=8cm]{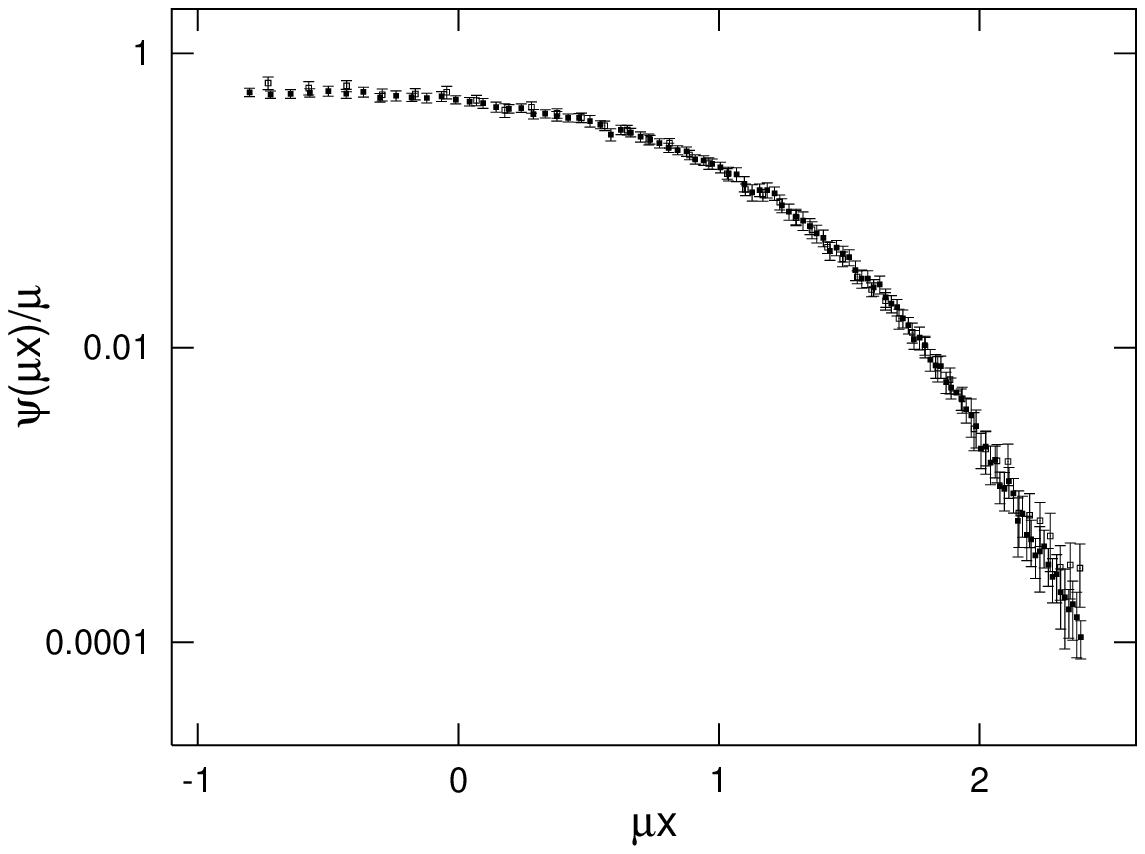}}
\hbox{\hspace{-.4cm}\includegraphics[width=8cm]{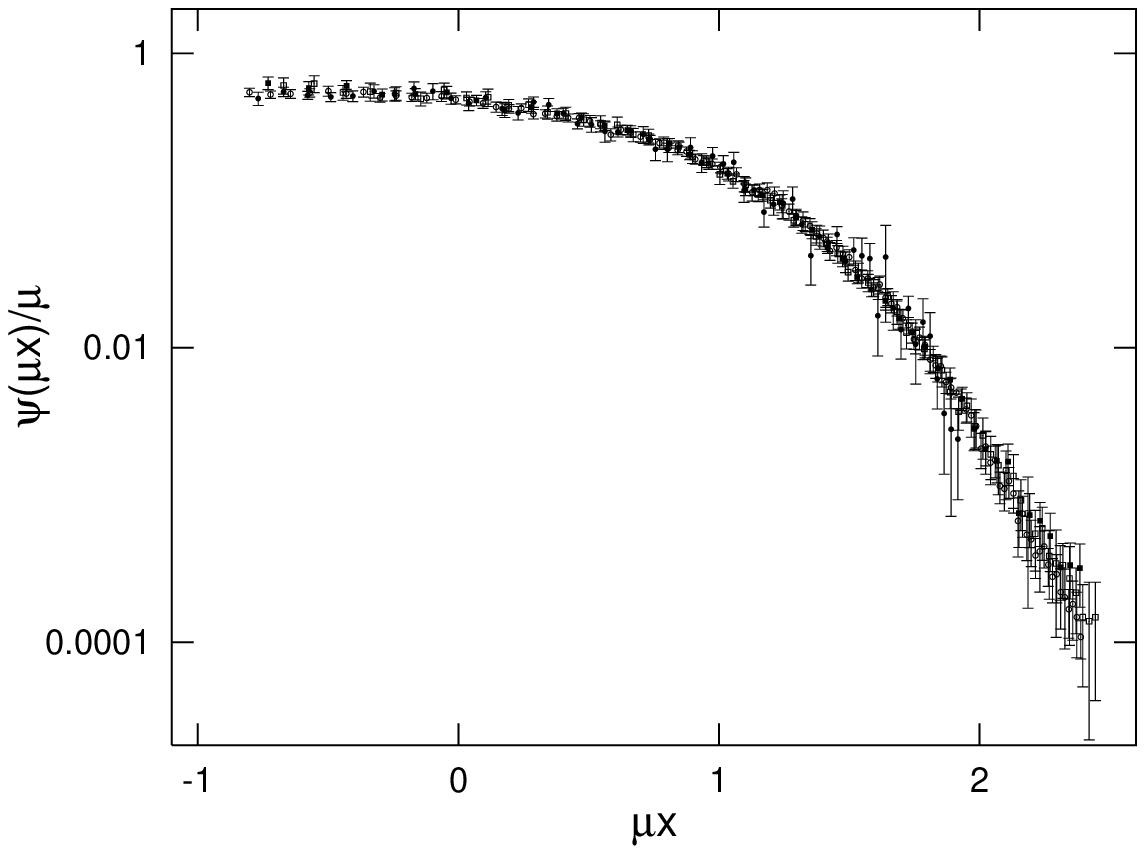}}
\end{center}

\vspace{-.3cm}

In the upper figure the $\sqrt s=$ 300 and 1800 GeV data sets 
are displayed only, 
whereas in the lower plot all the four data sets measured by the 
E735 experiment. The collapse of the data points onto a unique 
scaling curve is apparent. This very preliminary investigation suggests
that the pattern of KNO scaling violation by double parton collisions,
or by semihard processes dominated by minijet production, is similar
to that of $e^+e^-$ annihilations where log-KNO behavior is expected 
to arise as well at high energies (see before). 
But note that in parameter $k$ there 
is an order of magnitude difference between $e^+e^-$ and $p\bar p$ 
collisions, that is, the scaling functions have 
different shapes in the two reactions.

One of the earliest reviews~\cite{ole} of Polyakov's pioneering work
concluded with the following speculative note:
{\rm ``Could it be that as in the theory of critical phenomena, while the
basic physics (Hamiltonians) between $e^+e^-$ annihilation and hadronic
multiparticle production could be entirely different, multiplicity
scaling may yet still be a \ub{universal} feature shared by both?''\/}
In the light of our findings the answer to this question is {\it yes\/}
but the scaling behavior foreseen 30 years ago arises in the logarithm
of multiplicity instead of multiplicity itself.

\section{SUMMARY}

In the past 30 years, data collapsing of multiplicity distributions $P_n$
according to the generic KNO scaling relation Eq.~(2) was one of the most
extensively studied topics in soft physics. Unfortunately, more and more 
experimental and theoretical results confirm that Eq.~(2) is too simple 
to be true. The information content of $P_n$
can be represented in an equally convenient manner 
making use of the multiplicity moments.
For example, validity of Eq.~(2) corresponds to constancy of the normalized
central moments for appropriately chosen $c$ and $\lambda$. What can we do
if the KNO scaling rule~(2) breaks down? How data collapsing behavior of 
$P_n$ can be restored, if it is~possible at all? 
The approach presented here is extremely
simple: let's perform the original scaling transformation (shifting and
\hbox{rescaling}) in the multiplicity moments' rank, that is, in Mellin space, 
rather than in  multiplicity.

The functional relation Eq.~(3) tells how things transform 
at the level of $P_n$.
Translation in Mellin space (moment shifting) corresponds to changing from
$P_n$ to the moment distributions $P_{n,r}\propto n^rP_n$. 
This makes possible
to arrive at data collapsing for intermittency phenomena: 
e.g. $P_{n,1}$ scales
for monofractal fluctuations, $P_{n,2}$ for bifractals and so on. Dilatation 
in Mellin space (rescaling of moments' rank) corresponds 
to the transformation
$P_n\to P_m$ with $m=n^\mu$. Therefore data collapsing arises in rescaled 
logarithmic multiplicity (log-KNO). Such behavior is expected to occur for
phenomena governed by running coupling effects or in multiplicative cascades
of varying length. Considering the physical significance of the mentioned
examples in multiparticle dynamics, it will be interesting to see how the
proposed scaling behaviors show up in the experiments of the Third
Millennium.

\section*{ACKNOWLEDGEMENTS}

I am grateful to the organizers of this workshop, in particular
to Alberto and Roberto for the invitation and kind hospitality
in Torino. 

The support of grants NWO-OTKA N25186 and OTKA T024094 is acknowledged.

\end{document}